\documentclass[a4paper,twocolumn,english,prl,superscriptaddress,showpacs,longbibliography,fixfloat]{revtex4-1}
\usepackage[latin9]{inputenc}
\usepackage{xcolor}
\usepackage{pdfcolmk}
\usepackage{amsmath}
\usepackage{amssymb}
\usepackage{graphicx}
\PassOptionsToPackage{normalem}{ulem}
\usepackage{ulem}

\makeatletter

\pdfpageheight\paperheight
\pdfpagewidth\paperwidth

\providecolor{lyxadded}{rgb}{0,0,1}
\providecolor{lyxdeleted}{rgb}{1,0,0}
\usepackage[colorlinks,citecolor=darkblue,linkcolor=darkred,urlcolor=darkblue] {hyperref}
\usepackage{xcolor}
\definecolor{darkblue}{rgb}{0.1,0.2,0.6} 
\definecolor{darkred}{rgb}{0.8,0.1,0.2}
\renewcommand{\BibitemShut}[1]{}

\makeatother

\usepackage{babel}
\begin{document}

\title{Transport in quasiperiodic interacting systems: from superdiffusion
to subdiffusion}

\author{Yevgeny Bar Lev}

\affiliation{Department of Chemistry, Columbia University, 3000 Broadway, New
York, New York 10027, USA}
\email{yevgeny.barlev@columbia.edu}

\author{Dante M. Kennes}

\affiliation{Department of Physics, Columbia University, 3000 Broadway, New York,
New York 10027, USA}

\author{Christian Klöckner}

\affiliation{Dahlem Center for Complex Quantum Systems and Fachbereich Physik,
Freie Universität Berlin, 14195 Berlin, Germany}

\author{David R. Reichman}

\affiliation{Department of Chemistry, Columbia University, 3000 Broadway, New
York, New York 10027, USA}

\author{Christoph Karrasch}

\affiliation{Dahlem Center for Complex Quantum Systems and Fachbereich Physik,
Freie Universität Berlin, 14195 Berlin, Germany}
\begin{abstract}
Using a combination of numerically exact and renormalization-group
techniques we study the nonequilibrium transport of electrons in an
one-dimensional interacting system subject to a quasiperiodic potential.
For this purpose we calculate the growth of the mean-square displacement
as well as the melting of domain walls. While the system is nonintegrable
for all studied parameters, there is no\emph{ finite region} of parameters
for which we observe diffusive transport. In particular, our model
shows a rich dynamical behavior crossing over from superdiffusion
to subdiffusion. We discuss the implications of our results for the
general problem of many-body localization, with a particular emphasis
on the rare region Griffiths picture of subdiffusion.
\end{abstract}
\maketitle
\textit{Introduction. \textemdash{}} The finite energy transport properties
of quantum-mechanical systems generally fall into one of the three
standard categories: ballistic, diffusive, and the complete absence
of transport altogether. Ballistic transport is only possible in special,
so-called integrable cases (such as free fermions) where an extensive
set of local conserved quantities prevents currents from scattering.
Interacting, clean systems are generically diffusive, while the cessation
of transport is a hallmark of systems that completely fail to equilibrate,
such as those that undergo Anderson localization \cite{Anderson1958b}.
The study of the transition between such regimes has been energized
by the recent focus on systems that undergo many-body localization
(MBL) \cite{Basko2006a,Altman2014,Nandkishore2014}. In the standard
quenched disordered variants of such systems, dynamical behavior is
observed that exhibits a rich set of distinct transport behaviors
\cite{BarLev2014,Lev2014,Agarwal2014,Luitz2016c,Agarwal2016_review}.
Interacting quasiperiodic (QP) systems are also believed to exhibit
similar behavior, and are the basis for several recent experimental
studies on MBL \cite{Schreiber2015a,Luschen2016a}. Unfortunately,
very little is known about transport in such systems. Here, we fill
this vital gap via a finite-temperature version \cite{Karrasch2012a,Barthel2013,Kennes2014}
of the time-dependent matrix renormalization group (tDMRG) \cite{White1992,Schollwock2011}
as well as the functional renormalization group (FRG) \cite{Salmhofer1999,Metzner2012}.
\begin{figure}
\begin{centering}
\includegraphics{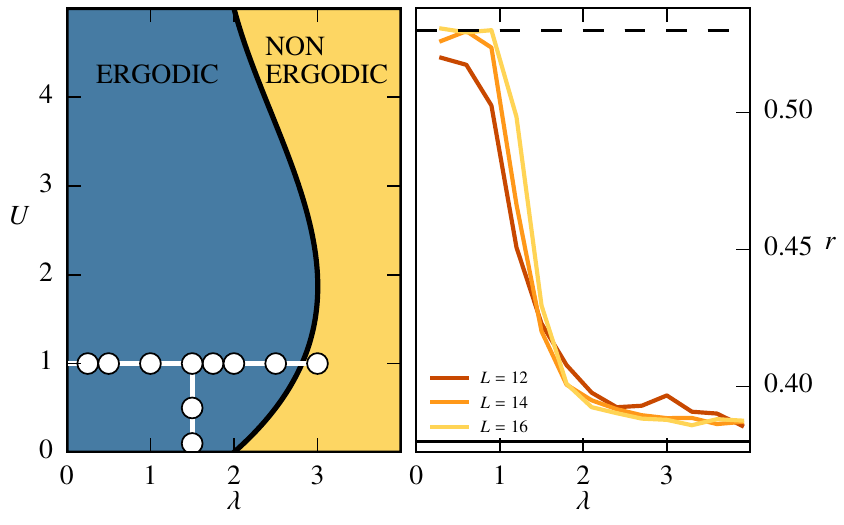}
\par\end{centering}
\caption{\label{fig:statics}(Left) A schematic phase diagram $\left(U,\lambda\right)$
of the system (c.f. Fig.~19 in Ref.~\cite{Bera2016a}). The white
circles represent the points studied in this work. (Right) The average
gap ratio $r$ in the middle of the spectrum for various strengths
of the QP potential, $U=1$, and different system sizes (lighter shades
represent larger system sizes). The dashed black line corresponds
to the Wigner-Dyson limit and the solid black line to the Poisson
limit. }
\end{figure}

\textit{Model. \textemdash{}} We consider a one-dimensional model
of spinless fermions, subject to a quasiperiodic potential (QP),

\begin{align}
\hat{H} & =\sum_{m=1}^{L-1}\left[J\left(\hat{c}_{m}^{\dagger}\hat{c}_{m+1}+\text{h.c.}\right)+U\hat{n}_{m}\hat{n}_{m+1}\right]+\sum_{m=1}^{L}V_{m}\hat{n}_{m},
\end{align}
which using the Jordan-Wigner transformation exactly maps to XXZ spin
model \cite{Jordan1928}. Here $L$ is the length of the system, $\hat{c}_{m}^{\dagger}$
$\left(\hat{c}_{m}\right)$ create (annihilate) a spinless fermion
on site $m$, ``h.c.'' denotes Hermitian conjugate, $\hat{n}_{m}$
is the corresponding density, $J$ is the hopping rate, which without
the loss of generality we set to $J=1$, $U$ is the interaction strength
and $V_{m}$ is a QP potential of the form,
\begin{equation}
V_{m}=\lambda\cos\left(2\pi\delta m+\phi\right),\label{eq:aa_potential}
\end{equation}
where $\delta$ is an irrational number which we will take to be inverse
of the golden mean, $\delta=\left(\sqrt{5}-1\right)/2$ and $\phi$
is a random phase. For $U=0$, this model reduces to the Aubry-André
model, which has a transition between all localized and all extended
single-particle states which occurs at $\lambda/J=2$ \cite{Aubry1980}.
For $\lambda>2J$ transport is absent in this model, while for $\lambda<2J$
transport is ballistic \cite{Kohmoto1989}. For $\lambda=0$ the Hamiltonian
is integrable even in the presence of interactions $\left(U\geq0\right)$.
For $U<2$ one can construct quasilocal conserved quantities to demonstrate
that transport is ballistic \cite{Prosen2011a,Ilievski2017}, while
for $U>2$ transport is diffusive \cite{Zotos1996,Steinigeweg2011,Znidaric2011,Karrasch2014}.
When both $\lambda$ and $U$ are nonzero the system is nonintegrable,
and has been studied numerically and experimentally. It has a MBL
transition which at infinite temperatures occurs for $\lambda/J>2$
\cite{Iyer2013,Schreiber2015a,Naldesi2016a}. We are interested to
study transport in regimes of weak integrability, in the limits where
the transport in the integrable limit is ballistic. Therefore through
this work we use $U\leq2$ (see left panel of Fig. ~\ref{fig:statics},
for the parameters we use). We would also like to avoid the MBL phase
where an emergent integrability appears \cite{Huse2013,Serbyn2013a,Ros2014,Imbrie2014,Imbrie2016,Imbrie2016a}.
For this purpose we search for the \emph{approximate} location of
the MBL transition in this model by calculating the gap ratio $r_{n}=\max\left(\delta_{n}/\delta_{n+1},\delta_{n+1}/\delta_{n}\right)$
in the middle of the spectrum, where $\delta_{n}$ are the eigenvalue
differences. The gap ratio is a convenient metric of short-range correlations
in the eigenvalues' statistics which was introduced in Ref.~\cite{Oganesyan2009}.
Ergodicity is assumed when the obtained probability distribution of
$r_{n}$ (or the unfolded $\delta_{n}$) matches the one of the corresponding
random matrix ensemble \cite{Atas2013}. In Fig.~\ref{fig:statics}
we calculate the mean value of $r_{n}$ averaged over 1000 realization
of the random phase in (\ref{eq:aa_potential}) and 50 values of $r_{n}$
computed in the middle of the many-body spectrum. We repeat this analysis
for $U=1$, a few system sizes, and various amplitudes of the QP potential,
and infer the location of the transition from the crossing point between
the different curves. Similarly to previous studies \cite{oganesyan_localization_2007},
the crossing point drifts to larger values of $\lambda$ when the
system size is increased, indicating that the transition occurs for
$\lambda>2$. This value is consistent with previous studies (c.f.
Fig.~19 in Ref.~\cite{Bera2016a}, but note a factor of 2 difference
in the definition of the kinetic energy). Since we endeavor in this
work \emph{not} to calculate the \emph{precise} location of the transition
(which is quite difficult to do, given the small system sizes which
are accessible) but to merely obtain the values of $\lambda$ for
which the system is clearly within the ergodic phase we focus mostly
on $\lambda\lesssim3$, although the extent of the ergodic phase might
be somewhat larger.

To characterize the transport we use a combination of a numerically
exact method, tDMRG and an approximate method, FRG, which allows us
to access significantly larger system sizes and longer times \cite{SuppMat2017}.
We evaluate the infinite temperature density-density correlation function,
\begin{equation}
C_{i,j}\left(t\right)=2^{-L}\text{Tr }\left(\hat{n}_{i}^{z}\left(t\right)-\frac{1}{2}\right)\left(\hat{n}_{j}^{z}-\frac{1}{2}\right),\label{eq:corr_func}
\end{equation}
and calculate the analog of the mean-square displacement (MSD),
\begin{equation}
x^{2}\left(t\right)=\frac{1}{L}\sum_{i,j=1}^{L}\left(i-j\right)^{2}C_{i,j}\left(t\right).\label{eq:msd}
\end{equation}
Normally the MSD scales as power-law with time, $x^{2}\sim t^{\alpha},$
where $\alpha$ is the dynamical exponent. For systems with ballistic
transport, $\alpha=2$, and for diffusive systems $\alpha=1$. Systems
with no transport or transport with a MSD growing more slowly than
any power law will have $\alpha=0$. To reduce the effects of the
boundaries we use systems of sizes $L=100-200$.
\begin{figure}
\begin{centering}
\includegraphics{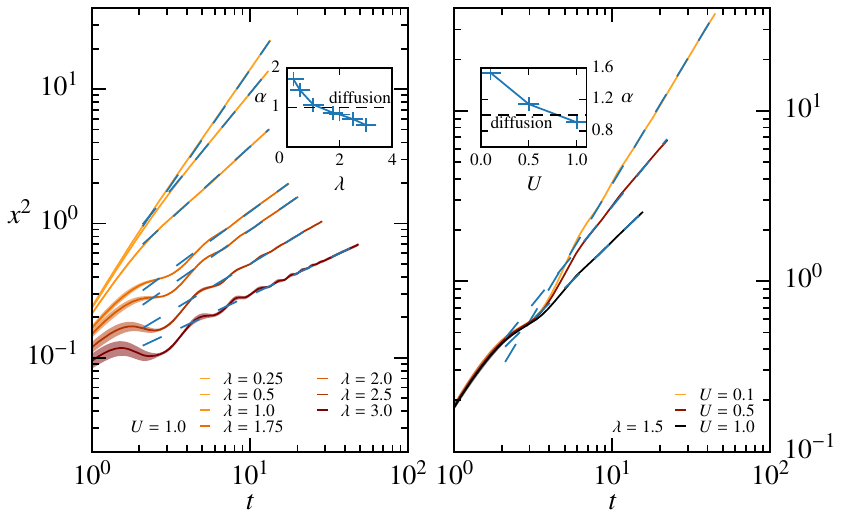}
\par\end{centering}
\caption{\label{fig:msd}Mean-square displacement as a function of time on
a log-log scale, for various amplitudes of the QP potential. The left
panel shows the horizontal cut through the phase-diagram at Fig.~\ref{fig:statics},
and the right panel shows the vertical cut. Darker colors represent
larger parameters, and the width of the lines represent the statistical
error bars. Blue dashed lines show the quality of the power-law fits,
and the insets present the corresponding dynamical exponents. The
system size used is $L=100$.}
\end{figure}

\textit{Superdiffusive regime. \textemdash{}} We start by analyzing
the vicinity of the integrable limits $U=0$ and $\lambda=0$. In
both cases, we do not observe a sharp crossover from ballistic to
diffusive behavior but uncover an extended \textit{superdiffusive
regime} where the MSD growth as a power law in time, $x^{2}\sim t^{\alpha}$,
with an exponent $1<\alpha<2$. This is illustrated in the left (right)
panels of Fig.~\ref{fig:msd} for the horizontal (vertical) cuts
through the phase diagram (see Fig.~\ref{fig:statics}). The occurrence
of superdiffusive transport in the presence of interactions and disorder
is in striking contrast to the behavior of clean systems where integrability
breaking terms normally lead to diffusion (for spin systems, see Refs.~\cite{Jung2006,Zotos2004,Steinigeweg2014,Karrasch2015b}).
However, a simple estimate of the mean free time of scattering from
the external potential gives a time-scale of $\tau\sim1/\lambda^{2}$,
which is about $\tau\approx16$ for the smallest $\lambda$ we study
and is comparable to our maximal simulation times. Therefore while
we convincingly observe superdiffusion over one decade in time, we
cannot rule out the scenario where it is merely a transient phenomenon.
Simulating for longer times is exponentially hard within tDMRG since
while it is a numerically exact method, the accessible time scales
are bounded by the growth of entanglement entropy. Therefore to substantiate
our observation of superdiffusion, we complement the tDMRG simulation
by a different approach which can reach to much longer times at the
price of being approximate. Since transport in the system is characterized
by power laws it is natural to use a renormalization-group based method
for this purpose.

\textit{Domain wall dynamics. \textemdash{}} In order to access longer
time scales and larger system sizes we use the functional renormalization
group (FRG) \cite{Salmhofer1999,Metzner2012} implemented on the real-time
Keldysh contour \cite{Jakobs2007,Kennes2012,Kennes2013}. The FRG
provides an \emph{a priori} exact reformulation of a quantum many-body
problem in terms of flow equations for vertex functions with an infrared
cutoff as the flow parameter \cite{SuppMat2017}. We truncate these
equations to leading order, which renders the framework approximate
with respect to the interaction strength. Due to its RG nature, the
FRG can capture power laws, and the corresponding exponents can be
computed up to the linear order in the interaction strength, $U$
(all higher-order contributions are uncontrolled). The computational
effort of the FRG calculation is not sensitive to the build up of
entanglement in the system, and scales linearly with time. In one
dimension one can access times of $t\sim1000$ for systems of up to
$L\sim1000$ sites. Since the MSD is a two-body correlation function,
it cannot be computed reliably using a first-order FRG scheme. To
circumvent this issue we investigate transport via a different protocol
which can be simulated both by the FRG and the tDMRG: the melting
of domain walls.
\begin{figure}
\includegraphics{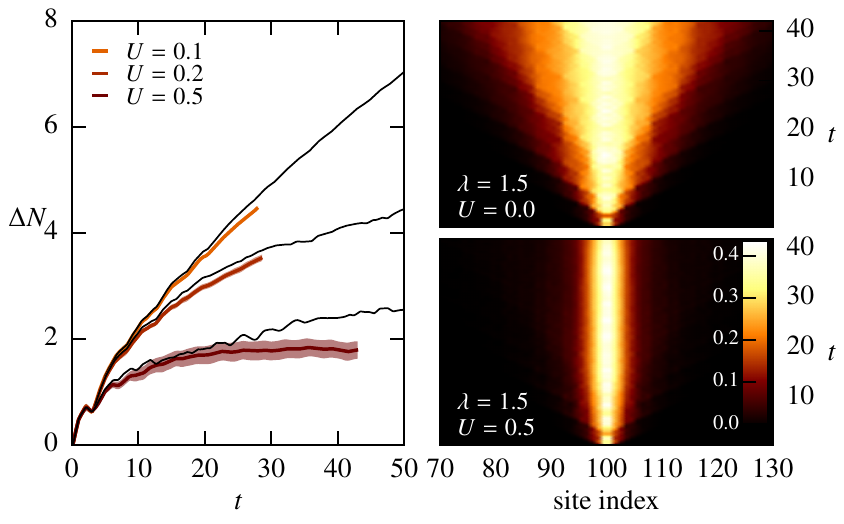}

\caption{\label{fig:domainwall}Mass transport for a domain wall initial condition.
The right side shows the mass difference, $\left|n_{i}\left(t\right)-n_{i}\left(0\right)\right|$,
as a function of time for two values of the interaction $U=0$ and
$0.5$, and demonstrates that transport, which is ballistic at $U=0$,
becomes successively suppressed when switching on interactions. The
left side compares tDMRG (colored lines) and FRG (black lines) results
for the total mass transferred to the left side as a function of time,
for fixed $\lambda$ and varying $U$. The width of the colored lines
represent the statistical error. }
\end{figure}

Domain wall dynamics provide a natural sensor for MBL physics \cite{Hauschild2016}
that can be realized straightforwardly in cold-atom experiments \cite{Choi2016}.
To be precise, we prepare the system in a state where all sites on
the left (right) are occupied (empty) and compute the number of particles
$\Delta N$ spreading between these two regions. In the localized
phase, the melting of the domain wall is suppressed, while in the
ergodic phase it is characterized by a power law growth of the transported
number of particles, $\Delta N\sim t^{\alpha/2}$, with the same exponent
$\alpha$, as the MSD (if calculated for the \emph{same} initial conditions).
We now study the ergodic phase close to integrability. Since FRG is
exact in the limit of $U=0$, we focus on the vertical cut through
the phase diagram in Fig.~\ref{fig:statics}. In the absence of interactions,
$\Delta N$ grows linearly with time indicating ballistic transport
(see right panel of Fig.~\ref{fig:domainwall}). Finite $U>0$ leads
to slower transport, and the FRG data is in good agreement with the
accurate reference provided by the tDMRG on intermediate time scales
(see left panel of Fig.~\ref{fig:domainwall}). 
\begin{figure}
\includegraphics{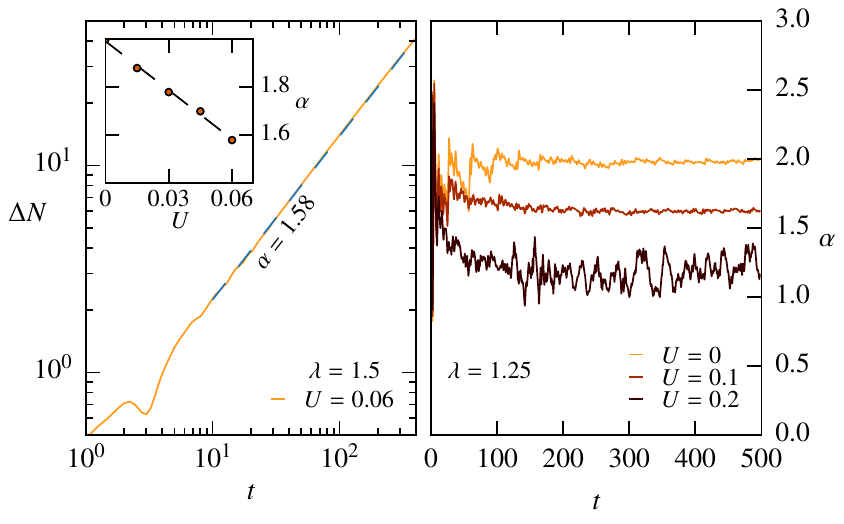}

\caption{\label{fig:log-deriv}(Left) Log-log plot of the transported number
of particles $\Delta N$ as a function of time. The blue dashed line
is a linear fit to the data from which the dynamical exponent is extracted,
and the inset shows the dynamical exponent as computed from FRG for
various interaction strengths, $\lambda=1.5$. (Right) Log-derivative
of the $\Delta N$ as function of time, $\lambda=1.25$. }
\end{figure}
The FRG can now be used to push the calculation to significantly larger
times, and a superdiffusive power law with $1<\alpha<2$ can be identified
unambiguously (see left panel of Fig.~\ref{fig:log-deriv}). Since
in this setup we use small interactions it is important to work with
times longer then the mean-free time of scattering between two particles,
$\tau_{ee}\sim1/U^{2},$ otherwise a transient ballistic transport
would be observed. Not only the transport we observe is always sub-ballistic,
but in our simulations we also use times which are about 2-20 times
longer then the scattering mean-free time $\tau_{ee}\sim25-277$ .
To verify that the extracted dynamical exponents do not drift with
time we also compute the log-derivative $\left(\mathrm{dlog}\Delta N/\mathrm{d}\log t\right)$
and observe that it saturates to a plateau (see right panel of Fig~\ref{fig:log-deriv}),
which indicates that the calculated dynamical exponents are asymptotic
within the FRG scheme. It is also important to check that the extracted
dynamical exponent $\alpha$ scales linearly with $U$, which is self-consistent
with the assumptions of FRG (see inset of Fig.~\ref{fig:log-deriv}). 

In the vicinity of $U=0$, the dynamical exponents agree qualitatively
with those governing the growth of the MSD (see Fig.~\ref{fig:msd}).
A strict quantitative comparison is however not possible since we
did not find a parameter set for which the exponents can be determined
reliably in \emph{both} setups. e.g., at $\lambda=1.5$, $U=0.1$
(see Fig.~\ref{fig:msd}), the domain-wall exponent is no longer
in the purely linear regime, and the higher-order corrections cannot
be computed reliably by the FRG. More importantly, there is no general
reason to expect that the exponents governing the spreading of the
domain wall and the MSD, computed for an infinite temperature initial
condition, should coincide: the former is a far out-of-equilibrium
initial condition while the latter describes linear response. The
difference between both setups becomes particularly apparent at larger
$U\sim0.5$ where we can no longer unambiguously identify power laws
in the evolution of the domain wall but observe a strong suppression
of transport. Since this effect appears to be a specific hallmark
of the (domain wall) initial conditions and does not directly coalesce
with the focus of this work, we leave the exploration of this interesting
regime to future work. We have thus provided evidence that transport
close of the integrable (ballistic) limits of vanishing interactions
or disorder is superdiffusive. We will now demonstrate that subdiffusion
materializes for larger amplitudes of the QP potential.

\textit{Subdiffusive regime. \textemdash{}} For interacting one-dimensional
systems subject to an uncorrelated disordered potential, transport
is surprisingly subdiffusive \cite{BarLev2014,Lev2014,Agarwal2014}
(see also recent reviews \cite{Luitz2016c,Agarwal2016_review} and
references within). It has been proposed that subdiffusion is a result
of rare spatial regions with anomalously large escape times (which
for example could correspond to areas with very short local localization
lengths) \cite{Agarwal2014,Gopalakrishnan2015,Gopalakrishnan2015a,Agarwal2016_review}.
These rare regions dramatically affect transport in one dimension,
since every particle has to pass through all effective barriers. This
picture, dubbed the \emph{Griffiths picture}, does not generically
apply when there are long correlations in the underlying disorder
potential, in particular in the case where the potential is QP \cite{Gopalakrishnan2015,Gopalakrishnan2015a}.
The Griffiths picture thus should predict diffusion for QP potentials
\cite{Gopalakrishnan2015}. 

\begin{figure}
\includegraphics{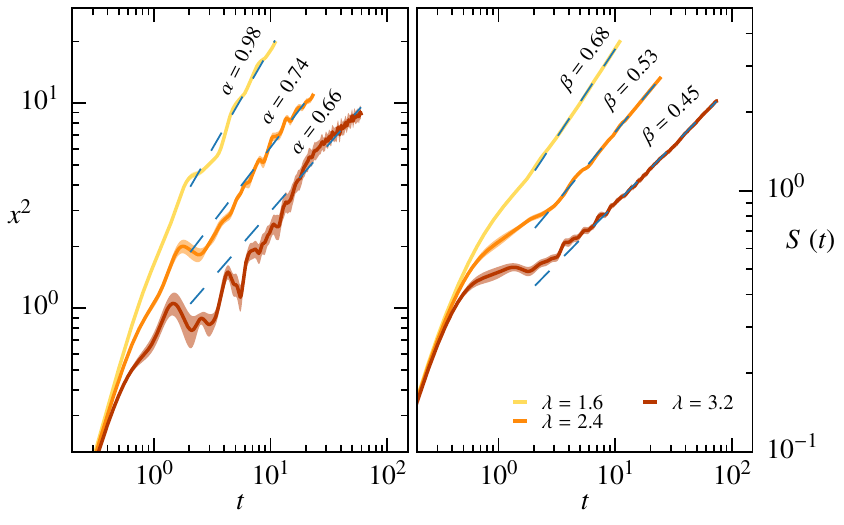}\caption{\label{fig:neel}Mean-square displacement (left) and entanglement
entropy (right) on a log-log scale as a function of time for the Néel
state as the initial state. Parameters used: $L=100$, interaction
strength $U=2$ and $\lambda=1.6,\,2.4$ and $,3.2$ (darker colors
designate higher values). Dashed lines represent linear fits used
to extract the dynamical exponents. Diffusion corresponds to, $\alpha=\beta=1$.}
\end{figure}

On the left panel of Fig.~\ref{fig:msd} we present the MSD as a
function of time for various strengths of the QP potential and $U=1.$
The computation was carried out at infinite temperature (cf. Eq.~(\ref{eq:corr_func})).
From the inset, which shows the extracted dynamical exponent, it is
clear that there is actually no \emph{finite} regime of parameters
for which the system is diffusive. Similar behavior was observed in
an experimental and numerical study, which appeared while this work
was in preparation \cite{Luschen2016a}. To verify that the observed
behavior occurs also for \emph{pure} initial states, we calculated
the MSD and the entanglement entropy (EE) starting form the Néel state
(see Fig.~\ref{fig:neel}). We note that for the system we study
the Néel state is a state with relatively high energy density, lying
close to the center of the many-body band, and has been successfully
utilized to demonstrate MBL in cold atoms experiments \cite{Schreiber2015a,Luschen2016a}.
However unlike the experiments, we do \emph{not} allow volatility
in the initial state, namely we have exactly one particle sitting
on every other lattice site. Similarly to the infinite temperature
initial state, for the Néel state initial condition both the MSD and
the EE show power law growth with time with dynamical exponents which
depend on the amplitude of the QP potential (EE was also studied in
Ref.~\cite{Naldesi2016a}). We note that while for the Néel state
initial condition the growth of the MSD appears to be subdiffusive
for the simulated times the extraction of the exponent is extremely
unreliable due to presence of oscillations in the data, \emph{which
do not disappear with better averaging}. This precludes from making
meaningful comparison between the dynamical exponent of the EE $\left(\beta\right)$
with the dynamical exponent of the MSD $\left(\alpha\right).$ For
a comparison of such exponents in disordered systems the interested
reader is referred to Ref.~\cite{Luitz2016c}.

\textit{Discussion. \textemdash{}} In this Letter we have demonstrated
that a simple but generic one dimensional interacting system with
a quasiperiodic potential exhibits an unexpectedly rich dynamical
behavior, exhibiting a crossover from superdiffusive to subdiffusive
transport. Within the study of systems connected to the problem of
many-body localization, the discovery of superdiffusive behavior is
a new one. While superdiffusion is expected to exist on \emph{finite}
time scales asymptotically close to an integrable point, our finding
that it also holds for substantial interaction and amplitude strengths
is surprising.

We have also presented numerical evidence which is inconsistent with
the prevailing explanation for subdiffusion in MBL systems, namely
the Griffiths picture \cite{Agarwal2014,Gopalakrishnan2015,Gopalakrishnan2015a,Agarwal2016_review}.
The Griffiths picture naturally relies on the presence of uncorrelated
quenched disorder in the system, which is crucial for generating a
sufficient density of rare blocking inclusions. Therefore for systems
with uncorrelated disorder in dimensions greater than one, those with
strongly correlated disorder, or for the quasiperiodic case studied
here, the Griffiths picture should yield asymptotic diffusive transport.
Contrary to these predictions we observe subdiffusive spin transport
and a sublinear spreading of entanglement entropy. We note that while
our numerical results are valid only for relatively short time scales,
there is no \emph{a priori} reason why subdiffusive behavior should
manifest within the Griffiths picture on any time scale in the quasiperiodic
case studied here. Our results are in line with a very recent experimental
study, which appeared while our work was in preparation \cite{Luschen2016a}.
There it was argued that while rare regions cannot be a result of
the quasiperiodic potential, they may follow from rare spatial regions
in the initial state \cite{Luschen2016a}. We stress that the infinite
temperature state, which we use as the initial state here, is clearly
translationally invariant, but still exhibits subdiffusion. Moreover
we have verified that subdiffusion is robust also when a \emph{pure}
initial state without any special spatial structure is taken (here
we considered the experimentally relevant Néel state).

Given our results, we speculate that subdiffusion is a result of atypical
\emph{transition rates} between the eigenstates of the \emph{noninteracting}
Hamiltonian and not necessary atypical \textit{spatial regions}. We
order the eigenstates of the noninteracting Hamiltonian according
to their energies and define the transition rates between any two
states, $\alpha$ and $\beta$ to be given by its Golden rule value,
$W_{\alpha\beta}=\left|V_{\alpha\beta}\right|^{2}/\left(E_{\alpha}-E_{\beta}\right)$,
where $V_{\alpha\beta}$ are the matrix elements of the Hamiltonian
terms which couple $\alpha$ and $\beta$. This picture is akin to
phenomenological trap models of glassy systems in configurational
space \cite{Monthus1996}. In particular, these models posit a \textquotedblleft particle,\textquotedblright{}
representing a location in parameter or phase space, which hops on
a random energy landscape and with broadly distributed waiting times.
Indeed given a master equation of the form 
\begin{equation}
\frac{\mathrm{d}P_{n}}{\mathrm{d}t}=W_{n,n-1}\left(P_{n-1}-P_{n}\right)+W_{n,n+1}\left(P_{n+1}-P_{n}\right),
\end{equation}
with a distribution of hopping times $p\left(W\right)\sim W^{-\alpha}$,
then subdiffusion with an exponent that smoothly decreases and vanishes
at a well-defined transition, as well as all of the scaling relations
normally associated with the Griffiths picture of MBL, are naturally
obtained \cite{Alexander1981,Luitz2016c}. It is important to emphasize
that the process we consider is not associated with the classical
exploration of a complex energy landscape by activation processes
through contact with the environment, but instead the transitions
are induced internally. Indeed, a recent calculation of the dynamics
of Anderson localization on the Bethe lattice, long believed to be
a proxy for MBL in low space dimensions, shows subdiffusive behavior
strikingly similar to that observed near the MBL transition and comports
with the trap-like model picture presented above \cite{Biroli2017}.

We would like to stress that while there is an apparent similarity
between the mechanism we propose and the Griffiths picture, in the
case of our mechanism the dependence on the dimensionality of the
physical system is quite weak, since any $d-$dimensional system is
mapped effectively to a model with no spatial structure. Moreover
there is no direct connection between the waiting time within a \textquotedblleft trap\textquotedblright{}
in the configurational space and \emph{spatially} atypical regions.
Therefore unlike the Griffiths scenario, our picture allows for subdiffusion
in higher dimensions as well as in the presence of long range spatial
correlations in the potential. Some evidence for subdiffusion in two
dimensions exists within the framework of self-consistent many-body
dynamics \cite{BarLev2015}. The microscopic mechanism of these atypical
transition rates, as well as a rigorous mapping to a quantum trap-like
model as envisioned above, have yet to be obtained and are certainly
goals worthy of future work. 
\begin{acknowledgments}
YB and DRR acknowledge funding from the Simons Foundation (\#454951,
David R. Reichman). We acknowledge support by the Deutsche Forschungsgemeinschaft
through the Emmy Noether program KA 3360/2-1 (CKl and CKa), as well
as through DK 2115/1-1 (D.K.). Part of the simulations in this work
were performed using computing resources granted by RWTH Aachen University
under project rwth0013 and rwth0057. YB, DK and CKl contributed equally
to this work.
\end{acknowledgments}

\bibliographystyle{apsrev4-1}
\bibliography{local,lib_yevgeny}

\end{document}